\documentclass[pteplogo]{ptephy_v1}

\preprintnumber{XXXX-XXXX} 
\usepackage{hyperref}

\usepackage[T1]{fontenc}
\usepackage{hyperref}
\usepackage{url}
\usepackage{siunitx}
\usepackage{multirow}
\usepackage[caption=false]{subfig}
\usepackage{svg}

\newcommand{\sqev}[1]{\SI{#1}{\electronvolt^2}}

\newcommand{\gev}[1]{\SI{#1}{\giga \electronvolt}}
\newcommand{\rad}[1]{\SI{#1}{\radian}}
\newcommand{\mrad}[1]{\SI{#1}{\milli \radian}}
\newcommand{\um}[1]{\SI{#1}{\micro \metre}}
\newcommand{\mm}[1]{\SI{#1}{\milli \metre}}
\newcommand{\cm}[1]{\SI{#1}{\centi \metre}}
\newcommand{\sqcm}[1]{\SI{#1}{\centi \metre^2}}
\newcommand{\km}[1]{\SI{#1}{\kilo \metre}}
\newcommand{\kt}[1]{\SI{#1}{\kilo \tonne}}

\begin{document}

\title{Updated constraints on sterile neutrino mixing in the OPERA experiment using a new $\nu_e$ identification method}

\author[1]{\collaborator{OPERA collaboration} N. Agafonova}
\author[2]{\mbox{A. Alexandrov}}
\author[3]{\mbox{A. Anokhina}}
\author[4]{\mbox{S. Aoki}}
\author[5]{\mbox{A. Ariga}}
\author[5,6]{\mbox{T. Ariga}}
\author[7]{\mbox{A. Bertolin}}
\author[8]{\mbox{C. Bozza}}
\author[7,9]{\mbox{R. Brugnera}}
\author[2]{\mbox{S. Buontempo}}
% ...

\dedi{To the memory of Prof. Pierre Vilain}

\affil[1]{INR—Institute for Nuclear Research of the Russian Academy of Sciences, RUS-117312 Moscow, Russia}
\affil[2]{INFN Sezione di Napoli, I-80126 Napoli, Italy}
\affil[3]{SINP MSU—Skobeltsyn Institute of Nuclear Physics, Lomonosov Moscow State University, RUS-119991 Moscow, Russia}
\affil[4]{Kobe University, J-657-8501 Kobe, Japan}
\affil[5]{Albert Einstein Center for Fundamental Physics, Laboratory for High Energy Physics (LHEP), University of Bern, CH-3012 Bern, Switzerland}
\affil[6]{Faculty of Arts and Science, Kyushu University, J-819-0395 Fukuoka, Japan}
\affil[7]{INFN Sezione di Padova, I-35131 Padova, Italy}
\affil[8]{Dipartimento di Fisica dell’Universit\'{a} di Salerno and “Gruppo Collegato” INFN, I-84084 Fisciano (Salerno), Italy}
\affil[9]{Dipartimento di Fisica e Astronomia dell’Universit\'{a} di Padova, I-35131 Padova, Italy}
% ...

\begin{abstract}

This paper describes a new $\nu_e$ identification method specifically designed to improve the low-energy ($< \gev{30}$) $\nu_e$ identification efficiency attained by enlarging the emulsion film scanning volume with the next generation emulsion readout system. A relative increase of 25-70\% in the $\nu_e$ low-energy region is expected, leading to improvements in the OPERA sensitivity to neutrino oscillations in the framework of the 3 + 1 model. The method is applied to a subset of data where the detection efficiency increase is expected to be more relevant, and one additional $\nu_e$ candidate is found. The analysis combined with the $\nu_\tau$ appearance results improves the upper limit on $\sin^2 2\theta_{\mu e}$ to 0.016 at 90\% C.L. in the MiniBooNE allowed region $\Delta m^2_{41} \sim \sqev{0.3}$.

\end{abstract}

\subjectindex{C04, C32}

\maketitle

\section{Introduction}

Oscillations among three neutrino flavours were established by solar, atmospheric, reactor and long-baseline accelerator neutrino experiments \cite{sno,sk,kamland,dayabay,t2k,minos,nova} over the last two decades. On the other hand, the presence of additional sterile neutrinos could explain the excess of $\nu_e$ and $\overline{\nu}_e$ charged-current events in short-baseline accelerator experiments---LSND \cite{lsnd} and MiniBooNE \cite{miniboone2018, miniboone2021}---and the deficits of $\nu_e$ and $\overline{\nu}_e$ from radioactive-source and reactor experiments \cite{ractor_anomaly, radioactive_source, best_experiment}.

The OPERA experiment was operated as a long-baseline neutrino oscillation experiment performed to observe the appearance of $\nu_\tau$ in a $\nu_\mu$ beam through the identification of their charged current (CC) interactions in a lead plates target instrumented with high resolution nuclear emulsion films \cite{opera_experiment}. The OPERA detector was exposed to the CERN Neutrinos to Gran Sasso (CNGS) $\nu_\mu$ beam \cite{opera_cngs} with a mean energy of about \gev{17} and was located at the LNGS underground laboratory, \km{732} away from the neutrino source. As a result of the data taking and the analysis, the OPERA Collaboration reported the discovery of $\nu_\tau$ appearance with a significance of 6.1$\sigma$ \cite{opera_tau_discovery, opera_final_nutau} and the results of a search for $\nu_e$ CC interactions in excess to expectation from the beam contamination \cite{opera_final_nue}. In the combined analysis of these two appearance modes \cite{opera_final_param}, a 90\% C.L. upper limit on $\sin^2 2 \theta_{\mu e} = 4 \left| U_{\mu 4} \right|^2 \left| U_{e4} \right|^2$ was set to 0.019 for $\Delta m^2_{41} > \sqev{0.1}$, and the MiniBooNE best fit values, $\Delta m^2_{41} = \sqev{0.043}$, $\sin^2 2\theta = 0.807$ \cite{miniboone2021}, were excluded.

Since the $\nu_\mu$ beam flux drops above \gev{30}, a high $\nu_e$ detection efficiency for the low energy ($< \gev{30}$) region is crucial for the $\nu_e$ appearance search. However, the $\nu_e$ detection method used in the previous analysis \cite{opera_final_nue} has efficiencies of 10-40\% for this energy region, because of the limited analysis capability due to the speed of emulsion readout systems at that time \cite{suts, ess}. Today, a 70 times faster scanning system makes it possible to improve the analysis \cite{hts}. 
In this paper we take advantage of this next generation system, to present a new $\nu_e$ identification method, report its performances and update the constraint on the parameters of the 3 + 1 neutrino mixing model.
\section{Detector, beam and data sample}
The OPERA detector was a hybrid apparatus made of nuclear emulsion trackers and electronic detectors \cite{opera_experiment}. The target was based on the Emulsion Cloud Chambers (ECCs) technology, consisting of alternating 57 emulsion films and 56 \mm{1} thick lead plates with a section of $12.7 \times \sqcm{10.2}$. The total thickness of \cm{7.5} was equivalent to about 10 radiation lengths. A pair of 2 films, called Changeable Sheet (CS) \cite{opera_experiment}, was attached externally on the downstream face of each ECC brick. The full detector had 2 identical super modules (SM), each of them was segmented into a target section and a muon spectrometer. In the target sections, ECC bricks were arranged in 29 layers of walls interleaved with target trackers (TT), which were planes of horizontal and vertical scintillator strips. A spectrometer, consisting of two iron core magnets instrumented with resistive plate chambers (RPCs) and drift tubes was mounted downstream of each instrumented target. It was aimed at identifying muons and measuring their charge sign and momentum. The two SMs contained about 150,000 ECC bricks corresponding to a total mass of \kt{1.25}.

The CNGS beam was an almost pure $\nu_\mu$ beam with a contamination of 2.0\% $\overline{\nu}_\mu$, 0.8\% $\nu_e$ and 0.05\% $\overline{\nu}_e$ \cite{opera_cngs_sim} in terms of CC interactions. The mean energy was \gev{17.9} ($\nu_\mu$), \gev{21.8} ($\overline{\nu}_\mu$), \gev{24.5} ($\nu_e$) and \gev{24.4} ($\overline{\nu}_e$). The prompt $\nu_\tau$ contamination was negligible $\mathcal{O}(10^{-7})$.

During the CNGS beam exposure of $17.97 \times 10^{19}$ protons on target, OPERA collected 19505 on-time events in the fiducial volume. The ECC brick where a neutrino interaction has occured was identified exploiting the pattern of the TT hits on-time with the CNGS beam. The track candidates in the CS are extrapolated to the ECC brick and searched upstream film by film. After location of the interaction vertex, $1 \times \sqcm{1}$ in 10 films downstream and 5 films upstream of the vertex were scanned and the tracks originated from the vertex were reconstructed. Finally 5868 neutrino interactions have been successfully reconstructed.
\section{New $\nu_e$ identification method}
\label{sec:method}

The events with one reconstructed muon or with a total number of fired TT and RPC planes larger than 19 were tagged as 1$\mu$ and excluded from the analysis \cite{opera_tau_discovery}. The remaining 1185 0$\mu$ events were targeted for the $\nu_e$ search.

The thickness of an ECC brick is enough to develop the electromagnetic (e.m.) showers produced by electrons originated from the $\nu_e$ CC interactions, whereas the scanning volume of \sqcm{1} and 10 films is equivalent to about 1.8 $X_0$. This scanning volume was limited by the scanning speed of the conventional readout systems \cite{suts, ess}. Therefore, in the previous search, identification of $\nu_e$ CC interactions was performed by a method using CS tracks, called CS Shower Hint (CSH). A search is performed in the CS for track segments less than 2 mm apart from the extrapolation point of each track originated from the interaction vertex (primary tracks). Moreover, the direction of candidate CS tracks is required to be compatible within \mrad{150} with that of the primary track. If at least three such tracks are found, additional scanning along the primary track is performed aiming at the detection of an e.m. shower \cite{opera_nue_2013}. 
However, the identification efficiency of $\nu_e$ CC events using the CSH method decreases with the $\nu_e$ energy because the probability for the e.m. shower to be absorbed in the ECC brick and not be able to reach the CS increases, especially if the interaction takes place in the most upstream part of the ECC brick. Improving the efficiency of detecting low-energy e.m. showers inside the ECC brick would increase the identification efficiency for low-energy $\nu_e$. To achieve this, a next generation emulsion readout system (Hyper-Track Selector, HTS) \cite{hts} with a scanning speed 70 times faster than conventional ones, has been introduced to enable enlargement of the scanning volume.

\begin{figure*}[!h]
    \centering
    \subfloat[Scanning volume and shower detection cone.]{
        \includegraphics[width=2.7in]{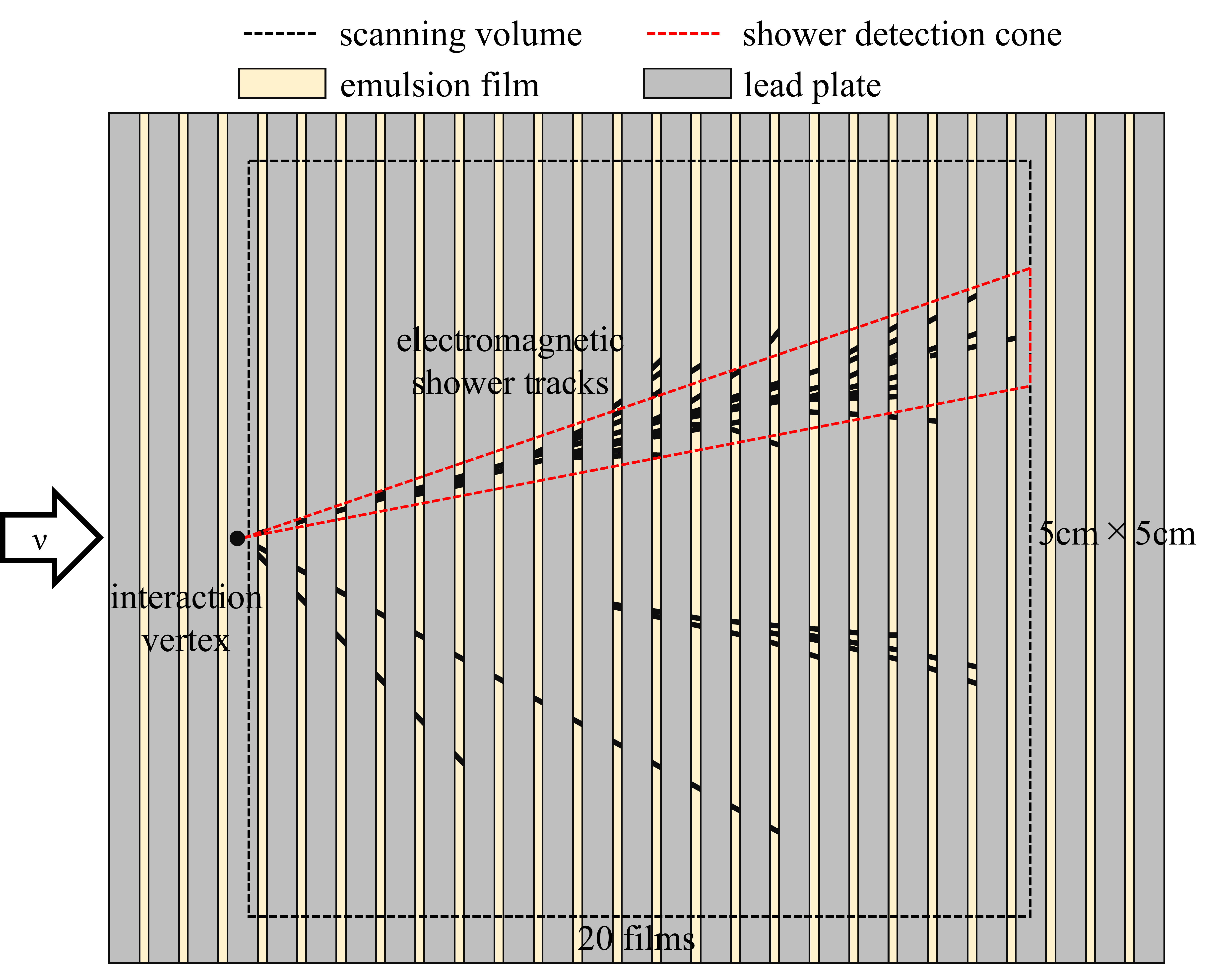}}
    \hfil
    \subfloat[Track segment selection criteria.]{
        \includegraphics[width=2.7in]{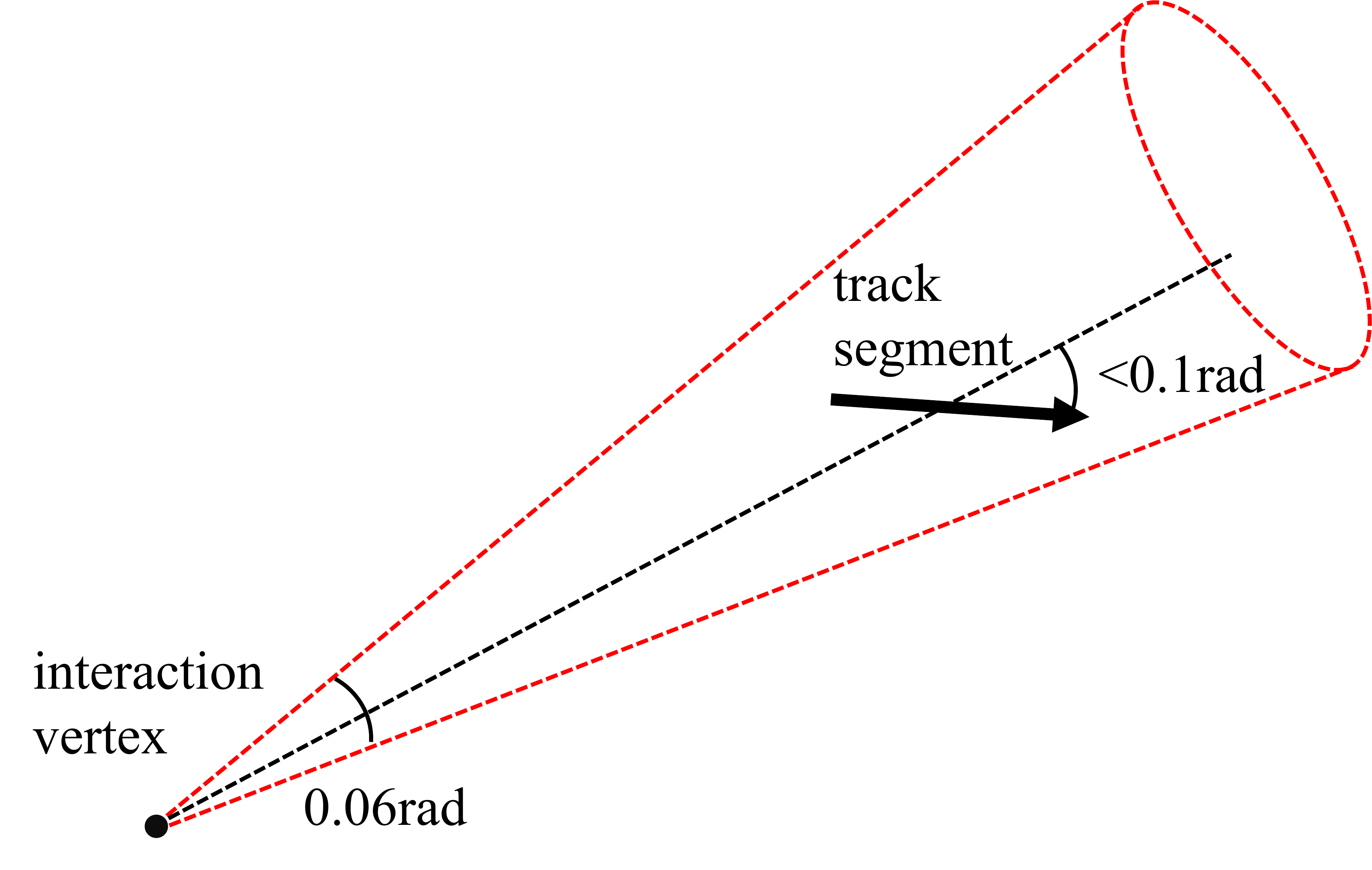}}
    \caption{The concept of the ECC Shower Detection (ESD) method}
    \label{esd_method}
\end{figure*}

The new $\nu_e$ identification method in ECC, hereafter called ECC Shower Detection (ESD), is defined as shown in Figure \ref{esd_method}. A volume of $5 \times \sqcm{5}$ and 20 films downstream from the vertex are scanned by HTS. After track reconstruction, the cones from the vertex with an apex angle of \rad{0.06} which contain at least 10 track segments pointing the vertex within a tolerance of 0.1 rad are tagged as e.m. shower candidates. These parameters have been optimised to keep a high detection efficiency while reducing noise. Tracks without showers and $e^+e^-$ pairs arising from at least 2 films downstream of the vertex are removed from the shower candidates by visual scan, while the other background sources are also removed by the selection described below. Remaining candidates are identified as $\nu_e$.

\begin{figure*}[!h]
    \centering
    \includegraphics[width=4.5in]{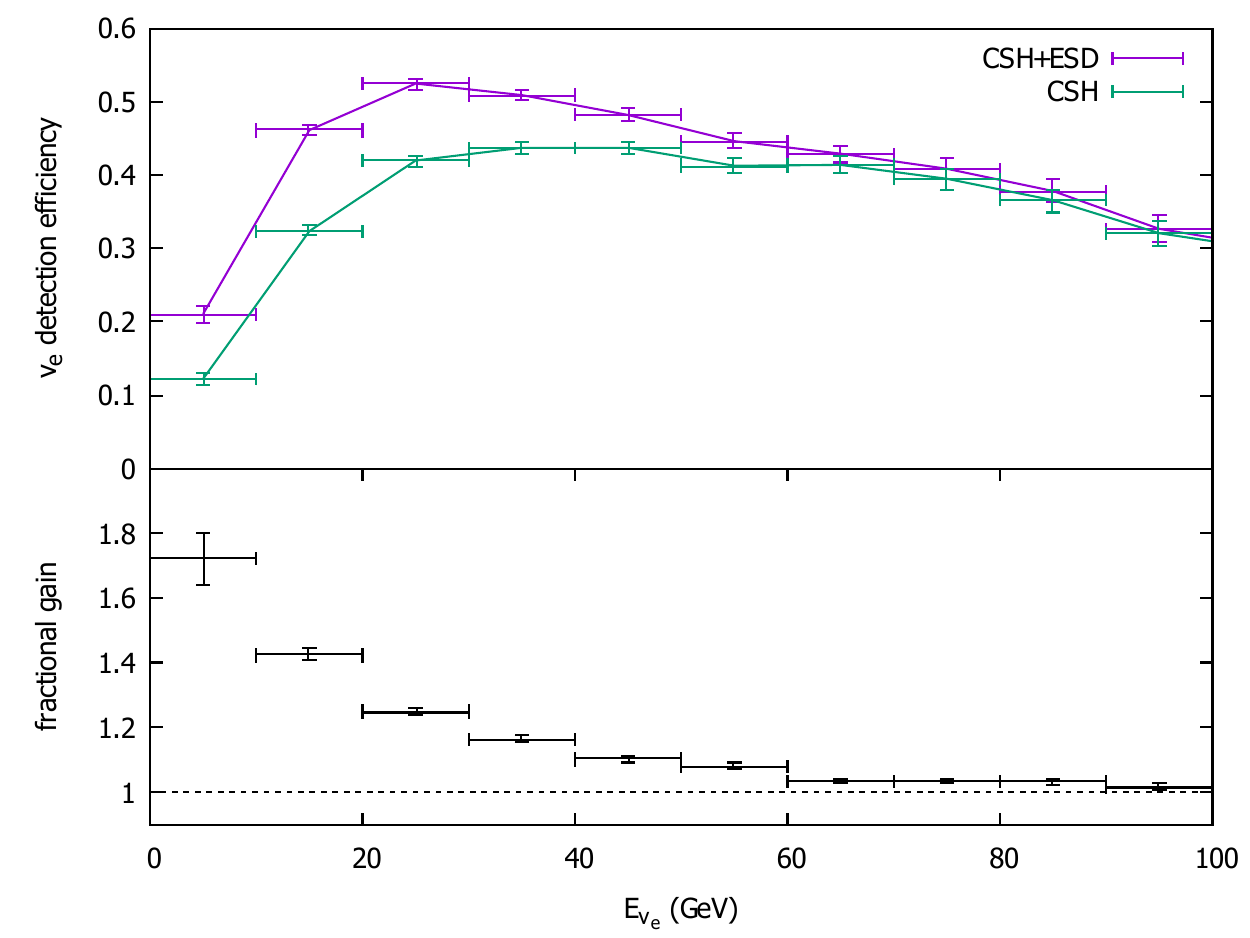}
    \caption{The top plot shows the $\nu_e$ detection efficiency and its statistical uncertainty as a function of the $\nu_e$ energy. The bottom plot shows the fractional gain of the efficiency, that is the ratio of the ESD+CSH efficiency and the CSH efficiency.}
    \label{efficiency_cs_vs_ecc}
\end{figure*}

The $\nu_e$ identification efficiency has been evaluated by a detailed Monte Carlo (MC) simulation with the standard OPERA simulation framework \cite{opera_sim_framework}, based on the CNGS beam fluxes estimated by FLUKA \cite{fluka1, fluka2} and the neutrino interactions generated by GENIE v2.8.6 \cite{genie1, genie2}. 
Here we describe the expected $\nu_e$ identification efficiency and number of $\nu_e$ CC interactions when analyzing the full data set, i.e. all 0$\mu$ events located in the ECC bricks. The detection efficiency of track segments, the position and angle accuracy, the track reconstruction and the shower detection process mentioned above with HTS are taken into account. The estimated $\nu_e$ identification efficiency is shown in Figure \ref{efficiency_cs_vs_ecc}; the improvement by adding the ESD method is about 25-70\% below \gev{30}. The expected number of $\nu_e$ CC candidates of the full data set without any other background in the no oscillation hypothesis increases from $31.0 \pm 0.9 \ (\mathrm{stat.}) \pm 3.0 \ (\mathrm{syst.})$ when using the CSH method only to $36.7 \pm 1.1 \ (\mathrm{stat.}) \pm 3.2 \ (\mathrm{syst.})$ when both the CSH and ESD methods are used. The improvement of the sensitivity for $\sin^2 2\theta_{\mu e}$ in the MiniBooNE allowed region $\Delta m^2_{41} \sim 0.3$ is expected to be 28\% by applying an energy spectrum analysis with the improved efficiency in the low energy region where it is more sensitive to the oscillations.

The ESD method has higher detection efficiency for low energy e.m. showers than the previous one, consequently it increases the backgrounds for the $\nu_e$ CC interaction search. The background sources are classified as follows: (1) $e^+ e^-$ pairs produced by prompt conversion of $\gamma$-rays from $\pi^0$ decays, (2) random coincidence of single hadron tracks and e.m. shower tracks, (3) $\tau \rightarrow e$ decays from $\nu_\tau$ CC interactions.

To limit/suppress the prompt $\gamma$ conversion background (1), a single electron track is searched for at the most upstream film in the scanning volume. The single electron classification requires that, in the region within \um{70} and \rad{0.3} from the primary electron candidate at the most upstream film, only hadron tracks are found or at most an even number of electron-like tracks. All tracks consisting of 8 or more track segments and no e.m. showers are classified as hadrons. An $e^+ e^-$ pair is misidentified as a single electron when one of the pair electrons is scattered out of the specified range for this criteria. In addition, the energy of the e.m. shower, estimated by a neural-network-based method using shower properties \cite{public_note_of_improved_nue_analysis}, is to be at least \gev{1.1}. These conditions have been determined to maximize the statistical significance of $\nu_e$ appearance.

Random coincidences of hadron and shower tracks (2) occur in 10\% of all 0$\mu$ events. They can be reduced by evaluating the probability that the primary electron candidate is a hadron. Thus a maximum likelihood estimation with the following 4 variables \cite{public_note_of_improved_nue_analysis} is applied: (a) minimum opening angle between the primary electron candidate and all $e^+ e^-$ pairs' direction, (b) ratio of the momentum of the primary electron candidate measured in the 1st-9th plates relative to the 10th-18th plates from the most upstream film, (c) number of films which primary electron candidate penetrates, (d) mean azimuthal opening angle between the primary electron candidate and the hadrons.

The conditions for $\tau \rightarrow e$ decays (3) to be identified as $\nu_\tau$ CC interactions are given in \cite{opera_final_nutau}. Conversely, they are wrongly identified as $\nu_e$ when they do not satisfy these conditions. After such selection, the expected numbers of backgrounds (1), (2) and (3) with their systematic uncertainty together with CSH are $2.6 \pm 0.9$, $1.2 \pm 0.5$ and $1.5 \pm 0.3$, respectively.
\section{Analysis sample} \label{scan_analysis}

\begin{figure}[!h]
    \centering
    \includegraphics[width=4.5in]{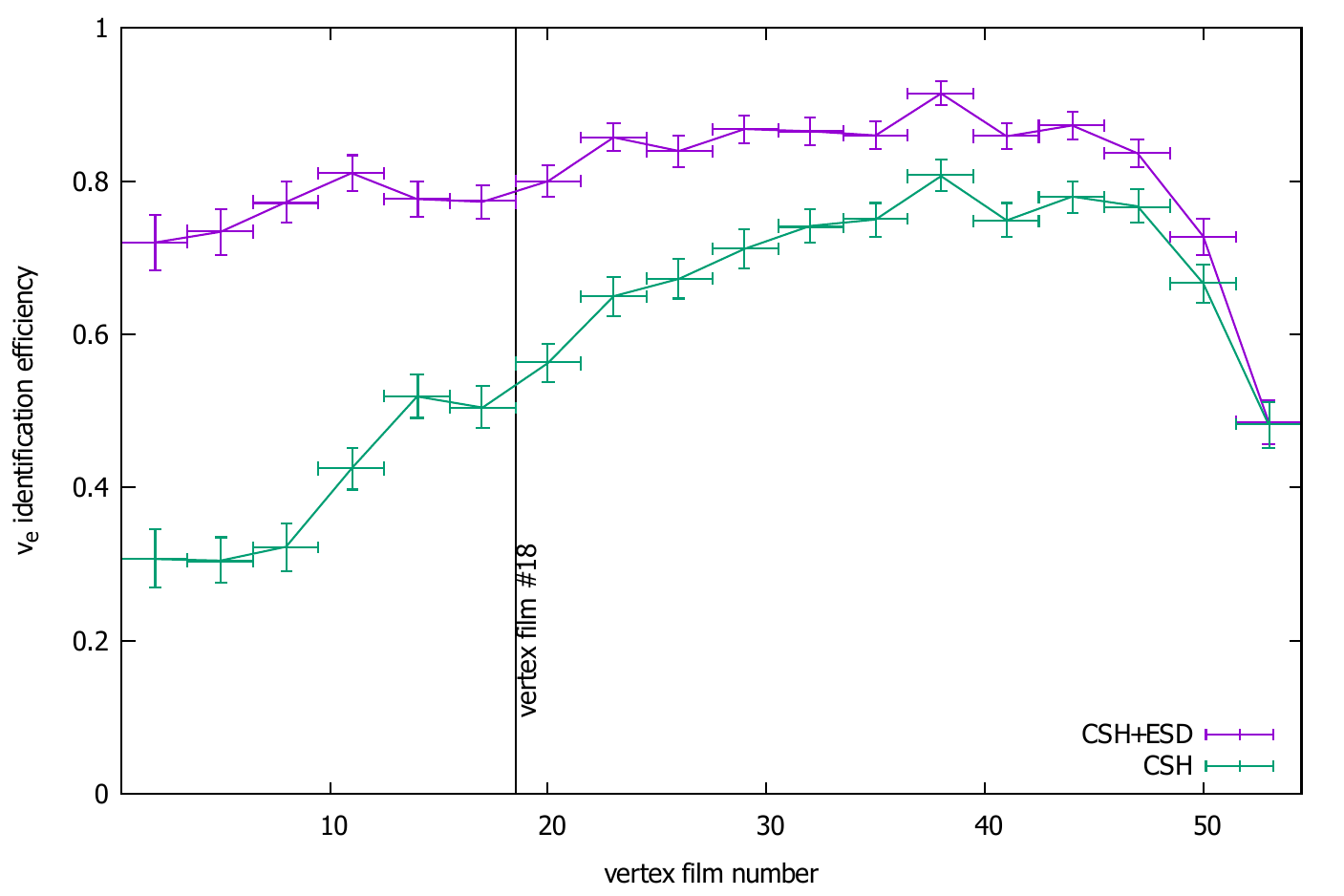}
    \caption{The $\nu_e$ identification efficiency as a function of vertex film number, for located events of energy $E_{\nu_e}<\gev{30}$. The beam enters the ECC brick through film \#1 and the CS is attached on the opposite side.}
    \label{nue_eff_vtx}
\end{figure}

In this section, we describe the result of the application of the ESD method to a sub-sample of 0$\mu$ events. In order to balance the requested additional scanning load and the expected number of additional $\nu_e$ candidates, we have selected a sub-sample according to the following criteria: the events have neutrino interactions in the upstream part of the ECC brick ($\mathrm{film \ number} < 18$, that is 7-10 $X_0$ deep from the downstream side of the ECC brick) and are contained within the volume of ECC brick at least \mm{5} away from the film edge, they show at least 2 reconstructed particles at the vertex and have not been identified as $\nu_e$ candidates by the CSH method. These criteria selected a sample of 99 events. Figure \ref{nue_eff_vtx} shows the $\nu_e$ identification efficiency as a function of the vertex film number (the number of the film immediately downstream of the vertex) for located events of energy $E_{\nu_e} < \gev{30}$, indicating that the efficiency improvement is higher for $\nu_e$ CC interactions located in the upstream part of the ECC brick.

The result of the scanning and the analysis is summarized in Table \ref{result_of_scan_and_analysis}. In the target events, 91 events were fully analyzed, while 8 were discarded due to failures in the analysis procedure. Track reconstruction failure is caused by the bad quality of the films and does not introduce biases, thus corresponding cases should be removed in the normalization procedure described below. On the contrary, the failure of vertex confirmation may produce a bias since it is likely associated to a lower track multiplicity at the primary vertex. However, the expected bias is estimated to be less than 1\%, smaller than the systematic uncertainty.

\begin{figure*}[!h]
    \centering
    \subfloat[3 flavour mixing.\label{spectra_3_osc}]{
        \includegraphics[width=2.7in]{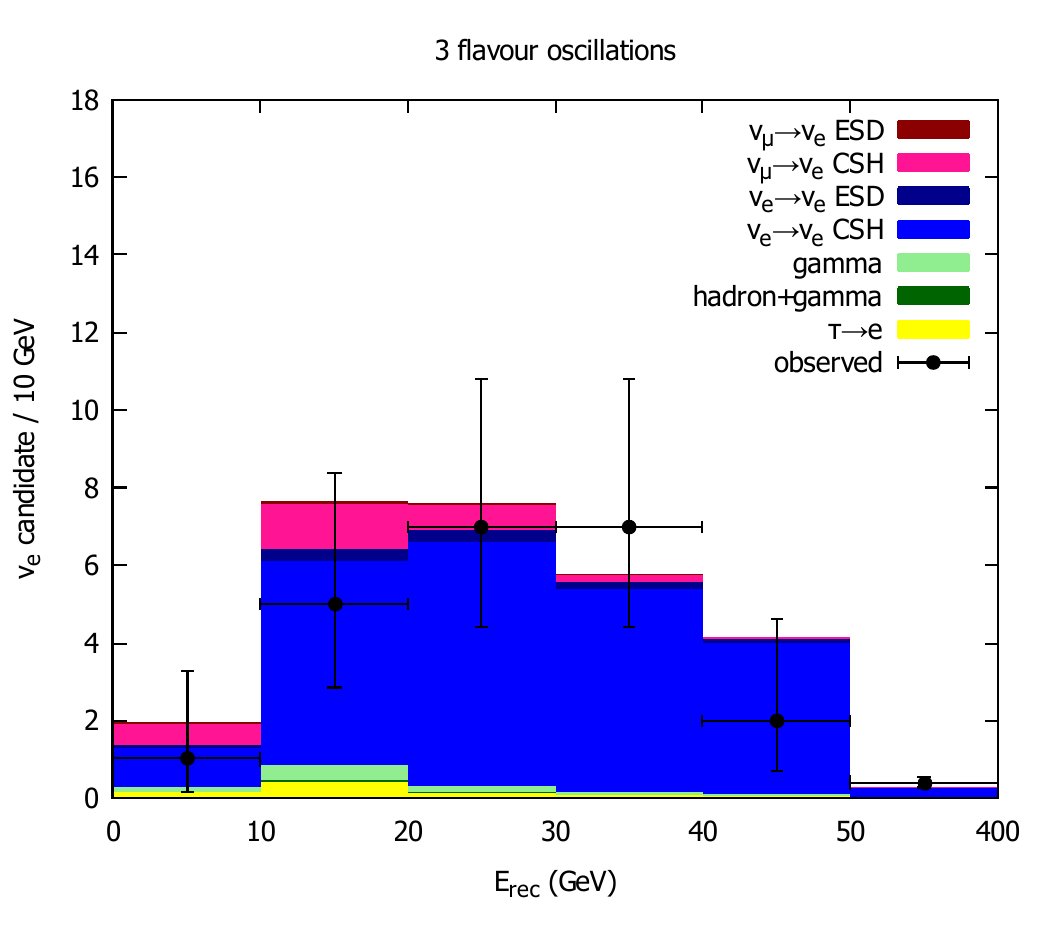}}
    \hfil
    \subfloat[3 + 1 flavour mixing.\label{spectra_3_1_osc}]{
        \includegraphics[width=2.7in]{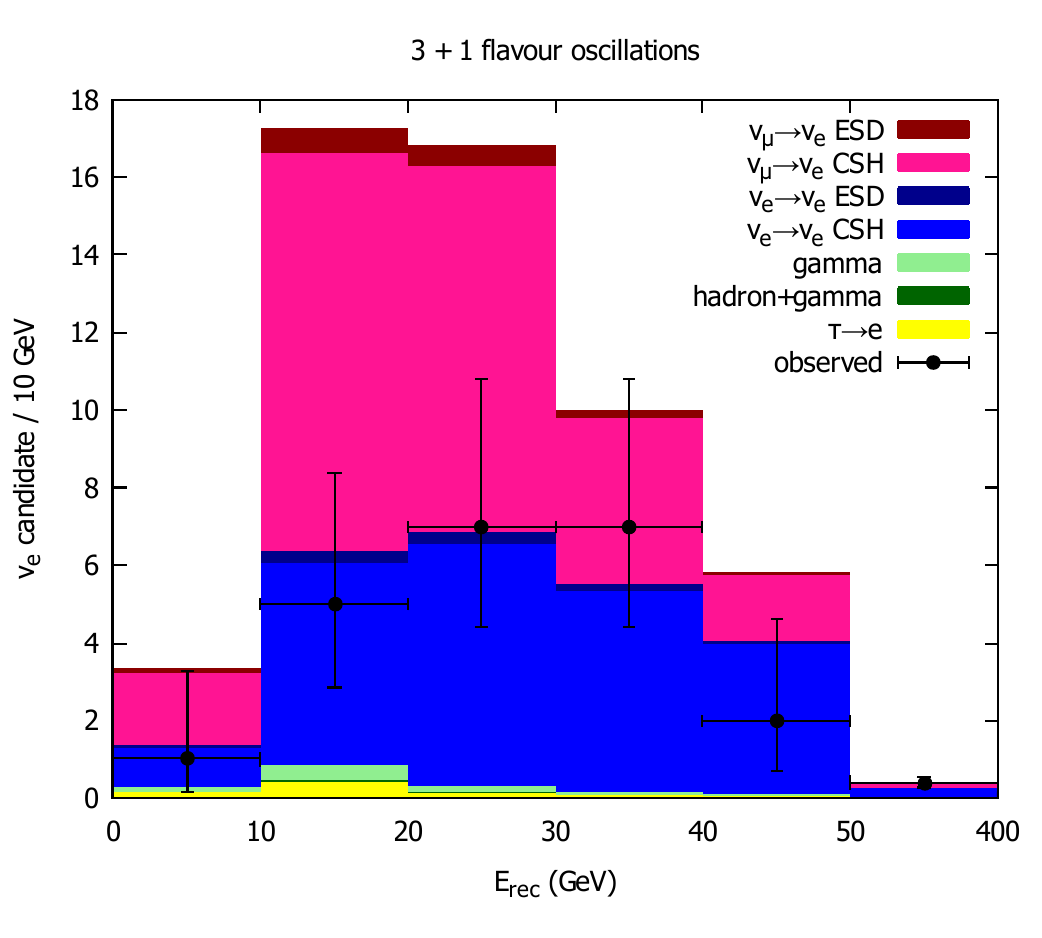}}
    \caption{Reconstructed energy distribution of expected and observed $\nu_e$ candidates on the assumption of \protect\subref{spectra_3_osc} the 3 flavour mixing and \protect\subref{spectra_3_1_osc} 3 + 1 flavour mixing with $\Delta m^2_{41} = \sqev{0.269}$, $\sin^2 2\theta_{\mu e} = 0.019$ and $P(\nu_e(\overline{\nu}_e) \rightarrow \nu_e(\overline{\nu}_e)) \simeq 1$.}
    \label{expected_and_observed_nue_candidates}
\end{figure*}

Electromagnetic showers were found in the analyzed volume of 48 events, and one of them was identified as a $\nu_e$ CC interaction candidate with a reconstructed energy of $(80 \pm 36) \, \si{\giga \electronvolt}$. 
A total of 14 events were observed in energy above \gev{50}, while the expected numbers of $\nu_e$ candidates from each source for this energy range are 0.04 ($\nu_\mu \rightarrow \nu_e$ ESD), 3.14 ($\nu_\mu \rightarrow \nu_e$ CSH), 0.11 ($\nu_e \rightarrow \nu_e$ ESD), 9.16 ($\nu_e \rightarrow \nu_e$ CSH) and 0.15 (other backgrounds) on the assumption of 3 + 1 mixing model with the same parameters as used in Figure \ref{spectra_3_1_osc}. 
The total number of observed $\nu_e$ candidates, including those detected by the CSH method, is 36.

\begin{table}[!h]
    \begin{center}
        \begin{tabular}{l|ll|r} \hline\hline
            & \multirow{2}{*}{e.m. shower found} & $\nu_e$ candidate & 1 \\
            Analysis completed & & $\gamma$ conversion & 47 \\
            & \multicolumn{2}{l|}{no e.m. showers} & 43 \\ \hline
            \multirow{2}{*}{Discarded} & \multicolumn{2}{l|}{track reconstruction failure} & 5 \\
            & \multicolumn{2}{l|}{vertex confirmation failure} & 3 \\ \hline
            \multicolumn{3}{l|}{Total} & 99 \\ \hline\hline
        \end{tabular}
    \end{center}
    \caption{Breakdown of the analysis result in terms of number of events.}
    \label{result_of_scan_and_analysis}
\end{table}

The expected number of prompt $\nu_e$ CC events in the absence of oscillation has been estimated using the CNGS flux, the neutrino cross section and the detection efficiency evaluated by the MC simulation incorporating the above-mentioned sub-sample selection. The normalization for the CSH method is described in \cite{opera_final_nue}, and for the ESD method the same normalization as CSH is applied with the fully analyzed 91 events. The systematic uncertainty on the expected number of events results from the combination of different uncertainties, part of which are common to both the CSH and ESD methods (CNGS flux, neutrino cross section and vertex location procedure), and the rest are specific (scanning and track reconstruction procedures including the visual scan by a human, MC sample statistics) and weighted by the ratio of the selected samples and the detection efficiencies between the two methods. The breakdown of them is shown in Table \ref{table_breakdown_of_syst_uncertainties_of_nue}. The contribution to the systematic error specific for the ESD method has been estimated to be 25\%/14\% for energies below/above \gev{10}, dominated by the limited size of the MC sample and the HTS track reconstruction efficiency. The overall systematic uncertainty is 19\%/10\% \cite{public_note_of_improved_nue_analysis}.

\begin{table}[htbp]
    \centering
    \begin{tabular}{l|l|rr}\hline\hline
        \multicolumn{2}{l|}{} & $<\gev{10}$ & $\geq \gev{10}$\\\hline
        \multicolumn{2}{l|}{Flux, cross section and location} & 14\% & 6\%\\\hline
        \multicolumn{2}{l|}{CSH identification} & 15\% & 8\%\\\hline
        & Track detection efficiency with HTS & 15\% & 9\%\\
        \multirow{2}{*}{ESD identification} & Difference of actual process with MC & 4\% & 3\%\\
        & Statistical uncertainty in MC & 20\% & 10\%\\
        & Overall in ESD identification & 25\% & 14\%\\\hline
        \multicolumn{2}{l|}{Overall} & 19\% & 10\%\\\hline\hline
    \end{tabular}
    \caption{Breakdown of the systematic uncertainties for the $\nu_e$ detection efficiencies of the CSH and ESD methods.}
    \label{table_breakdown_of_syst_uncertainties_of_nue}
\end{table}

Other background sources such as prompt $\gamma$ conversions, random coincidences between hadron and e.m. shower tracks and $\tau \rightarrow e$ decays have been estimated by the same MC simulation, using the same normalization as for the $\nu_e$ CC events (Table \ref{exp_from_each_source}). The combined systematic uncertainty of these background sources is dominated by the limited MC sample statistics and estimated to be 36\% \cite{public_note_of_improved_nue_analysis}.

The expected numbers of $\nu_e$ candidates from the beam incorporating the sub-sample selection are summarized in Table \ref{exp_from_each_source}. It should be noted that the total number of expected $\nu_e$ candidates by the ESD method with energy $<\gev{30}$ is $1.0 \pm 0.2$. The decreases in the expectations of the $\gamma$ and hadron + $\gamma$ from the numbers with the full data set are greater than that of the beam $\nu_e$. One of the major reasons is that the vertex film distributions of the $\gamma$-ray backgrounds are quite different from the beam $\nu_e$ because of the small e.m. shower energies, and another is that the $\gamma$-ray backgrounds have large statistical errors in the MC simulation.

\begin{table}[!h]
    \centering
    \begin{tabular}{l|cccc} \hline\hline
        & beam $\nu_e$ & hadron + $\gamma$ & $\gamma$ & $\tau \rightarrow e$ \\ \hline
        CSH & $31.0 \pm 3.0$ & negligible & $0.5 \pm 0.5$ & $0.7 \pm 0.2$ \\
        ESD & $1.1 \pm 0.1$ & $0.1 \pm 0.1$ & $0.3 \pm 0.2$ & $0.05 \pm 0.01$ \\ \hline\hline
    \end{tabular}
    \caption{The expected number of $\nu_e$ candidates from each source under the assumption of no oscillations and with the CSH and ESD methods applied to the sub-samples.}
    \label{exp_from_each_source}
\end{table}

The reconstructed energy distribution of expected and observed $\nu_e$ candidates are shown in Figure \ref{expected_and_observed_nue_candidates}. The oscillation parameters from \cite{review_of_pdg} are used for the 3 flavour mixing model. For the 3 + 1 mixing model, parameters on the intersection between the MiniBooNE allowed region \cite{miniboone2021} and the OPERA exclusion border \cite{opera_final_param}, i.e. $\Delta m^2_{41} = \sqev{0.269}$, $\sin^2 2\theta_{\mu e} = 0.019$, and $P(\nu_e(\overline{\nu}_e) \rightarrow \nu_e(\overline{\nu}_e)) \simeq 1$ are assumed. The total expected number of $\nu_e$ candidates from prompt $\nu_e$, $\overline{\nu}_e$ and all other backgrounds is 34.0 $\pm$ 3.3.

\begin{figure*}[!h]
    \centering
    \subfloat[]{
        \includegraphics[width=2.7in]{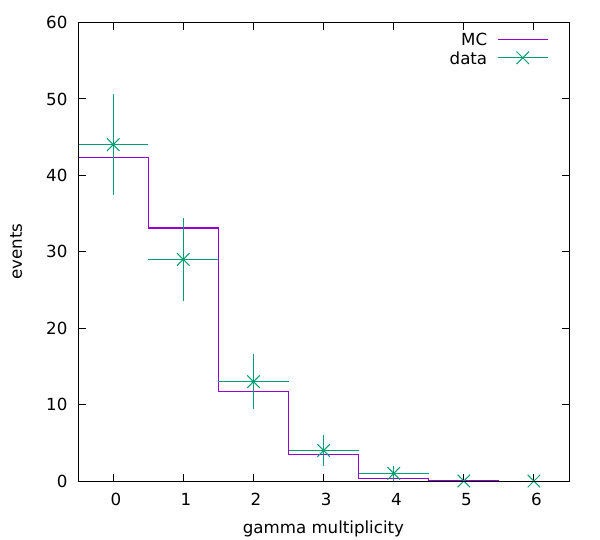}
        \label{gamma_multiplicity_distribution}}
    \subfloat[]{
        \includegraphics[width=2.7in]{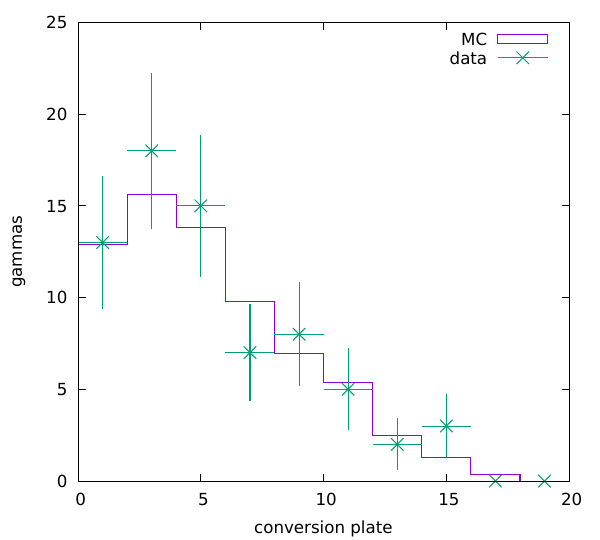}
        \label{distribution_of_gamma_free_path_before_conversion}}
    \caption{$\gamma$-ray multiplicity (left) and conversion plate number from vertex (right).}
    \label{gamma_ms_vs_data}
\end{figure*}

As already mentioned, the ESD method has a high sensitivity for low energy e.m. showers, therefore the comparison of the observed $\gamma$-ray properties to the expectation is useful for the validation of this method. In Figure \ref{gamma_ms_vs_data}, the distribution of the $\gamma$-rays multiplicity---the number of $\gamma$-rays detected per event---and their free path before conversion obtained from the MC simulation are compared to those of the observed $\gamma$-rays. The number of events in the MC simulation is normalized with the number of fully analyzed events. Both MC distributions are well consistent with experimental data, and the expected number of $\gamma$-rays, $69 \pm 11$, is in agreement with the observed one, 71.

\section{Oscillation analysis in the 3 + 1 mixing model}
In order to check the presence of a light sterile neutrino as suggested by the LSND and MiniBooNE experimental results, the 3 + 1 flavour mixing model is assumed. Not only $\nu_e$ but also $\nu_\tau$ appearance searches were conducted by the OPERA collaboration \cite{opera_final_nutau}. Since the $\nu_\tau$ flux with the parameters used in Figure 4b is expected to vary from almost 0 to 10 times larger than the 3 flavour mixing, and some parameter space allowed by only $\nu_e$ appearance analysis can be excluded, both appearance channels have been jointly used. 
The statistical analysis is based on the profile likelihood ratio by comparing the observed energy spectrum to the expectation under the 3 + 1 flavour mixing model. $\Delta m_{41}^2$ and $\sin^2 2\theta_{\mu e}$ are the parameters of interest. The value of $\Delta m^2_{21}$ is fixed to the PDG value \cite{review_of_pdg}, while a gaussian constraint on $\Delta m^2_{31}$ is assumed with mean and sigma also found in \cite{review_of_pdg}. The dependencies on the other parameters are removed by treating them as nuisance parameters. More details are described in \cite{opera_final_param}.

\begin{figure*}[!h]
    \centering
    \includegraphics[width=5in]{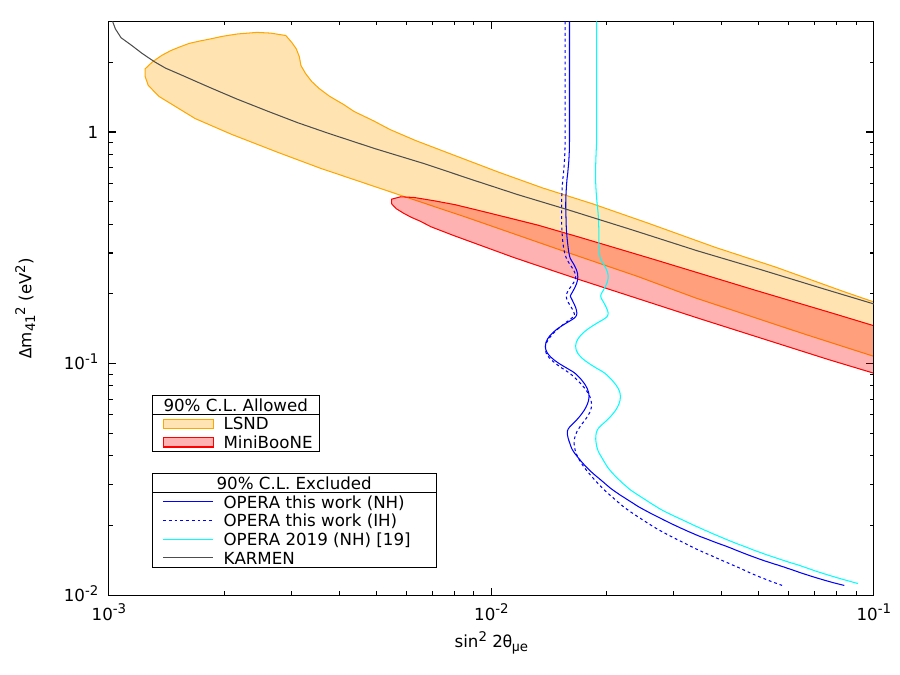}
    \caption{The 90\% C.L. exclusion region in the $\Delta m^2_{41}$ and $\sin^2 2\theta_{\mu e}$ plane for the normal (NH, solid) and inverted (IH, dashed) hierarchies. The allowed region at 90\% C.L. by LSND \cite{lsnd} and MiniBooNE \cite{miniboone2021}, the excluded region by the previous OPERA analysis \cite{opera_final_param} and KARMEN \cite{karmen} are also shown. The region drawn above is completely excluded by the combined result of MINOS, MINOS+, Daya Bay and Bugey-3 \cite{minos_dayabay_bugey3}, which conducted $\nu_\mu$ and $\nu_e$ disappearance analyses.}
    \label{exclusion_plot}
\end{figure*}

As the result of this test statistic, the 90\% C.L. exclusion region obtained by using both the CSH and ESD methods is shown in Figure \ref{exclusion_plot}. The upper limit around the MiniBooNE allowed region $\Delta m^2_{41} \sim \sqev{0.3}$ has been lowered to $\sin^2 2 \theta_{\mu e} < 0.016$.

\section{Conclusions}
A new $\nu_e$ identification method, called ESD, was introduced to improve the detection efficiency for low energy $\nu_e$ events that are crucial to investigate the MiniBooNE allowed region in the 3 + 1 mixing model. The shower detection method was optimised to detect electron-induced showers, and the $\nu_e$ identification efficiency increased by up to 70\% below \gev{30}.

We applied the method to a subsample of 99 0$\mu$ events with a vertex in the most upstream part of the ECC brick. We have found 1 new $\nu_e$ candidate with a reconstructed neutrino energy of $(80 \pm 36) \, \si{\giga \electronvolt}$. The expected additional number of $\nu_e$ candidates is $1.5 \pm 0.2$, in particular $1.0 \pm 0.2$ for energies below \gev{30}. It is worth noting that the observed $\gamma$-rays from $\pi^0$ decays are in agreement in number and properties with the expectation.

The 3 + 1 flavour mixing model has been tested and the 90\% C.L. constraints have been improved to $\sin^2 2\theta_{\mu e} < 0.016$ around the MiniBooNE allowed region $\Delta m_{41}^2 \sim \sqev{0.3}$.

\section*{Acknowledgment}
We warmly thank CERN for the successful operation of the CNGS facility and INFN for the continuous support given by hosting the experiment in its LNGS laboratory. Funding is gratefully acknowledged from national agencies and Institutions supporting us, namely: Fonds de la Recherche Scientifique-FNRS and Institut Interuniversitaire des Sciences Nucl\'{e}aires for Belgium; MZO for Croatia; CNRS and IN2P3 for France; BMBF for Germany; INFN for Italy; JSPS, MEXT, the QFPUGlobal COE program of Nagoya University, and Promotion and Mutual Aid Corporation for Private Schools of Japan for Japan; SNF, the University of Bern and ETH Zurich for Switzerland; the Programs of the Presidium of the Russian Academy of Sciences (Neutrino Physics and Experimental and Theoretical Researches of Fundamental Interactions), and the Ministry of Education and Science of the Russian Federation for Russia; the National Research Foundation of Korea (NRF) grant funded by the Korea government (MSIT) (No. 2021R1A2C2011003) for Korea; and TUBITAK, the Scientific and Technological Research Council of Turkey for Turkey (Grant No. 108T324).
\let\doi\relax

\end{document}